# Validating Digital Traces with Survey Data

The Use Case of Religiosity

M.Fuat Kına*
Sociology
Koç University

Erdem Yörük
Sociology
Koç University

Zübeyir Nişancı
Sociology
Marmara University

Ali Hürriyetoğlu
KNAW Humanities Cluster
DHLab

Fırat Duruşan
Center for Computational Social Sciences
Koç University

Osman Mutlu
Center for Computational Social Sciences
Koç University

Melih Can Yardı
Computational Social Sciences
Koç University

Oğuz Gürerk
Center for Computational Social Sciences
Koç University

Gizem Bacaksızlar Turbic
GESIS

Şükrü Atsızelti
Sociology
Koç University

Tolga Etgü
Mathematics
Koç University

Yusuf Akbulut
Sociology
Marmara University

**ABSTRACT**

This paper tests the validity of a digital trace database (Politus) obtained from Twitter, with a recently conducted representative social survey, focusing on the use case of religiosity in Turkey. Religiosity scores in the research are extracted using supervised machine learning under the Politus project. The validation analysis depends on two steps. First, we compare the performances of two alternative tweet-to-user transformation strategies, and second, test for the impact of resampling via the MRP technique. Estimates of the Politus are examined at both aggregate and region-level. The results are intriguing for future research on measuring public opinion via social media data.

## 1. Introduction

Big data from social media and AI methods offer an immense potential to measure public opinion and various trends in society, including voting behavior, consumer behavior, ideologies, perceptions, values, or beliefs (Barberá et al. 2015; Barberá 2015). Yet, both digital traces from social media and the AI methods used to analyze them carry a wide array of biases and limitations, which culminate in different types of errors. Therefore, the results of social media-based public opinion analysis need validation to identify and measure the errors. The lack of ground-truth data through which such validation could be conducted is a big challenge, especially when it comes to the analysis of public opinion, for which official data is intrinsically absent. Therefore, traditional opinion polls remain the only viable source to deliver validation data, but it is challenging to access or implement such surveys on particular topics of interest. In this paper, we present the results of a validation exercise, which is an outcome of a rather fortunate collaboration between two teams, working separately on big data and traditional survey data. Our results illustrate that social media offers a very promising opportunity to measure public opinion.

## 2. The Politus Project

The Politus project develops a computational social science approach to automatically extract public opinion in Turkey. The project creates a data platform that delivers representative, high-frequency, multilingual, multi-country, and privacy-protected panel data on key political and social trends in Turkey and selects other countries in later stages. By aggregating and automatically analyzing the digital trace data of social media users, the Politus platform makes population-level projections on public opinions—such as the prevalence of certain political ideologies, beliefs, and values, approval/support ratings of governments and political entities—and political behavior—such as voting preferences and trends. Said projections are geolocated, thus allowing local, as well as national, level analyses, and disaggregated according to key demographic variables, such as age, gender, ethnicity and race, religion, and education level. The project collects data initially from Twitter and processes it with ethically compliant deep learning models and natural language processing (NLP) tools (approved by the ERC Ethics Review and Expert Management Unit).

---

* Corresponding author, contact email:
mkina18@ku.edu.tr

A gold standard corpus of tweets, annotated by graduate students in social and political sciences and adjudicated by a domain expert, was used for the training of language models. Tweets presented to the annotators were randomly sampled and self-contained (i.e., did not include media or quotes from other posts). Annotation task classified tweets according to their subjects into politics and policy-related topics, and ideologies and beliefs expressed in them. Tweets that are labeled religious, which is one of the belief categories, were defined as containing any favorable statement of one's religious devotion and/or worshiping practice, citations of holy scripture and/or sunnah, preferring religious expressions and gestures in the language (i.e., while congratulating, showing appreciation, thanking, greeting, etc.). A transformer-based pre-trained language model (Reimers 2019) with a final linear layer is trained further using these annotations.[2] The model was trained to decide whether a tweet is religious or not and it achieves 92,21% F1 macro on the test set.

The project collects and stores a comprehensive database of Twitter users in Turkey. It collected user information from the followers of the most popular 100 accounts in Turkey, which corresponds to 53 million users. Then, it identified 3.5 million users, whose gender and location characteristics are determined via a public tool. 1 million users have age and gender predictions coming from the M3-Inference demographic inference tool[3] and 781K users have both these predictions and at least one tweet.

## 3. Validation Methodology and Results

In order to validate the religiosity estimation conducted by Politus, we analytically compare and contrast regional variations of religiosity scores with a recently implemented nationally and regionally representative survey, namely the Faith and Religiosity Survey of Turkey (Türkiye İnanç ve Dindarlık Araştırması, TIDA), which was conducted between December 2021 and May 2022 by a project led by co-authors. The survey data was collected from 1,942 people aged 18 and over, using random sampling methods, in twelve regions across Turkey at the Statistical Region Classification Level 1 (NUTS-1), and covering rural and urban populations proportionally. The TIDA focuses on different dimensions of religious beliefs, attitudes, and behaviors according to gender, age group, region, and education level. The main objective of the TIDA Project was to understand and analyze the distribution of common beliefs and religious practices across Turkey providing a more comprehensive framework in content and using more consistent techniques, compared to previous studies.

For the validation analysis, this research first acknowledges and measures the level of demographic bias in the Politus. The unrepresentative nature of digital traces is often considered one of the most challenging measurement errors while studying social media data, since participation in online networks is strongly affected by users' age, gender, education level, race, income level, etc. (Sloan 2017; Cohen and Ruths 2013; Filho et al. 2015; Barbera 2016; Olteanu et al. 2019; Sen et al. 2020). The demographic bias for the user-level Politus database is presented in Figure 1, which illustrates the distribution of observations for gender categories and age groups for the administrational data (taken from the Turkish Statistical Institute, TÜİK), the TIDA, and the Politus. Accordingly, females, older generations, and people living in underpopulated regions are underrepresented in the Politus database, which may rightfully cast doubt on its validity, especially for the gender gap. Males are observed approximately three times more present on Twitter than females. To tackle the problem of unrepresentativeness in Twitter, we resort to resampling the results via the Multilevel Regression and Poststratification technique (known as MRP or Mister P.). The MRP is based on the adjustment of every possible combination of characteristics according to their actual presence in the population through multilevel regression analysis (Lax and Phillips 2009; Park, Gelman, and Bafumi 2004). Our MRP models, therefore, employ the characteristics of gender, age, and region of residence to adjust the aggregate and region-level estimations in the research. We follow the workflow and publicly available R script of Leemann and Wasserfallen (2020) for the MRP analysis.

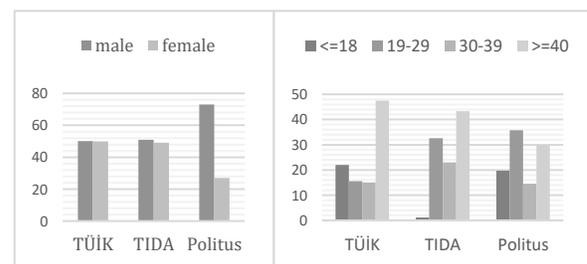

**Figure 1: Demographic bias in the Politus: Comparison of age groups (left), gender (right)**

---

[2] Accessed on https://github.com/politusanalytics/twitter_piousness_classifier, 09 November 2022

[3] Accessed on https://github.com/euagendas/m3inference. 17 August 2022

The validation process depends on two different metrics of religiosity that are extracted from the Politus database. The first one is a binary variable that represents whether the user is a religious person (*religious_dummy*), and the second is a continuous score of religiosity ranging from 0 to 1 (*religiosity_score*). Accordingly, we followed two strategies to transform tweet-level measures into user-level variables. The tweet-to-user transformation in this research is based on five tweet-level measures: the user's original, retweeted, favorited, quote tweets, and replies.[4] For the binary religiosity variable, we coded the user as religious if at least one tweet from any of the five categories above is predicted as a religious tweet in a dichotomous manner—with the default threshold of 0.5. For the religiosity score, however, we took the arithmetic average of the predicted probabilities of the tweets from all categories. We filtered user-level data with a corresponding conditionality: all users in the database have at least one tweet from any of the five categories in the last twenty-four months, which levels down the total sample size to 301,291 unique Twitter users.

The binary religiosity variable is positive for 47 percent of users in the Politus database. However, the distribution of predicted probabilities for the tweets is highly right-skewed with a mean value of 0.035 (out of 1.00). This seems to be an outcome of the fact that individuals are relatively reluctant to share their opinions when not asked, compared to a survey scenario, which is a platform affordance and reveals another challenging measurement error in this research (Sen et al. 2019). Therefore, we measure the magnitude of the gap between the tweet-level estimated religiosity scores and survey responses and concentrate on the regional variation for the validation analysis of the religiosity score, instead of the exact magnitudes. This difference between the two tweet-to-user transformation strategies also inspired us to compare the results of validation analysis for those, to see which one outperforms.

In the TIDA database, we focus on self-perceived religiosity, instead of its practical applications since the former conceptually better aligns with the self-declaring nature of tweeting. The survey question for self-perceived religiosity is asked as follows: "*How religious do you see yourself in general?*", and has five options: "*very religious*", "*religious*", "*neither religious nor*", "*not religious*", and "*not religious at all*". Correspondingly to the two tweet-to-user transformation strategies, we revised the structure of this variable. Like the *religious_dummy* in the Politus, we coded "*religious*" and "*very religious*" categories as positive and others as negative values. And we rescaled the original ordinal variable between 0 and 1 for the continuous religiosity score in the Politus. The descriptive statistics for all are presented in Table 1.

**Table 1: Descriptive statistics**

|  | Obs. | Mean | Std. | Min | Max |
|---|---|---|---|---|---|
| *religious_dummy_TIDA* | 1893 | 0.62 | 0.49 | 0 | 1 |
| *religiosity_score_TIDA* | 1893 | 0.62 | 0.25 | 0 | 1 |
| *religious_dummy_Politus* | 301291 | 0.47 | 0.50 | 0 | 1 |
| *religiosity_score_Politus* | 301291 | 0.04 | 0.06 | 0.01 | 1 |

To obtain comparable units between the TIDA and the Politus, we collapsed the datasets into the categories of four age groups, two gender, and twelve regions. Extracting ninety-six rows, this operation enables us to compare the average religiosity scores and the ratio of religious persons in TIDA and Politus throughout the subdivisions of the demographics (e.g., female, older than 40, living in Istanbul). To better interpret and compare the magnitudes of the correlation coefficients we also included two additional variables from the TIDA—that consider practical applications of religiosity as Ramadan fasting and praying—and present a Spearman correlation heatmap of all variables according to the ninety-six subdivisions in Figure 2.[5]

Comparing the coefficients in Figure 2, the Politus metrics are positively correlated with both the practical applications of religiosity and self-perceived religiosity measures in the TIDA, to varying magnitudes from 0.49 to 0.59.[6] Two alternative metrics of Politus are similar in their correlation coefficients. More importantly, the magnitudes of correlation coefficients are very close to the internal correlations of the TIDA variables (correlations between self-perceived religiosity measures and practical applications of religiosity). Therefore, potentially, the Politus metrics seem to bring us to the right track of capturing the internal variability among different religiosity metrics of the TIDA. Nevertheless, the religious ratio variable in the Politus is slightly superior, compared to the religiosity score variable.

---

[4] Note that the five tweet level measures in the research do not include reference tweets in replying and quotation, since these may not refer to an endorsement.

[5] We preferred Spearman correlation since it is more convenient for the ordinal data.
[6] All correlation coefficients in the table are statistically significant at 0.001 p value.

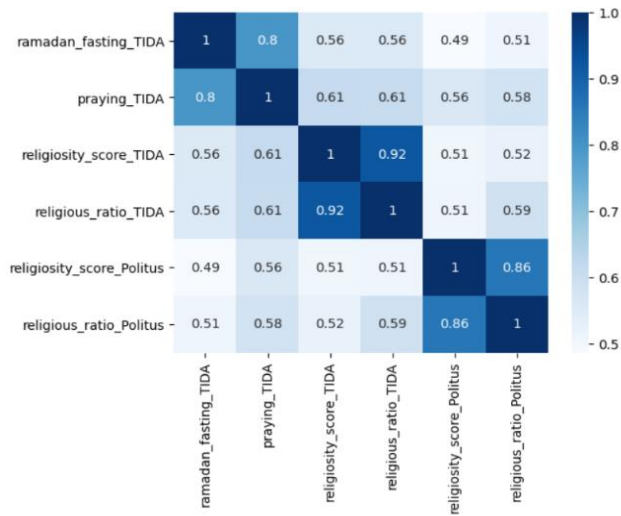

**Figure 2: Correlations of the Politus metrics and the TIDA variables**

However, in order to decide which metric in the Politus has a higher quality to capture the regional variation in the TIDA, we went one step further. Figure 3 and Figure 4 show the same results respectively on bar graphs and x-y coordinates, comparing disaggregated and post-MRP estimation results for the religious ratio, and bar graphs for the religious score, through the twelve regions of Turkey. Both the religiosity score and religious ratio variables of the TIDA serve as reference measures in the figures. While the country-level average scores for the reference data are 0.62 for the two variables in the TIDA, as shown in Table 1, the MRP analysis increases the country-level average score for the religious ratio metric of the Politus from 0.47 to 0.51. However, the results imply that each strategy of tweet-to-user transformation has a comparative advantage. Although the religiosity score is quite disadvantageous in magnitude, it outperforms in estimating regional variation (note that we revised the reference legend on the y-axis for the religiosity score in the Politus).

## 4. Conclusion

This research aimed to validate the religiosity estimates of the Politus database via the TIDA survey. Although the Politus metrics capture the regional variation and internal variability of the TIDA variables to a significant extent, the magnitudes of individual scores need improvement. Accordingly, we detected two potential sources of measurement errors for the Politus database that contain useful insights for other research. These errors correspond to the two limitations, which are at the same time the primary next steps of the Politus project. The first one was the demographic bias, for which we adjusted Politus estimates via the MRP analysis, and it provided little benefit. However, our MRP models did not employ a Bayesian framework, since this would bring out a high computational and temporal cost for our large N data. The second one was the default small magnitudes of the predicted probabilities. Nevertheless, the tweet-to-user transformation strategies in this research do not consider networks of Twitter users. Examining followers-followings information might significantly increase and adjust the Politus estimates. Researchers working with big data increasingly need to test the validity of their models, meaning that the exchange between social surveys and big data will become much more important.

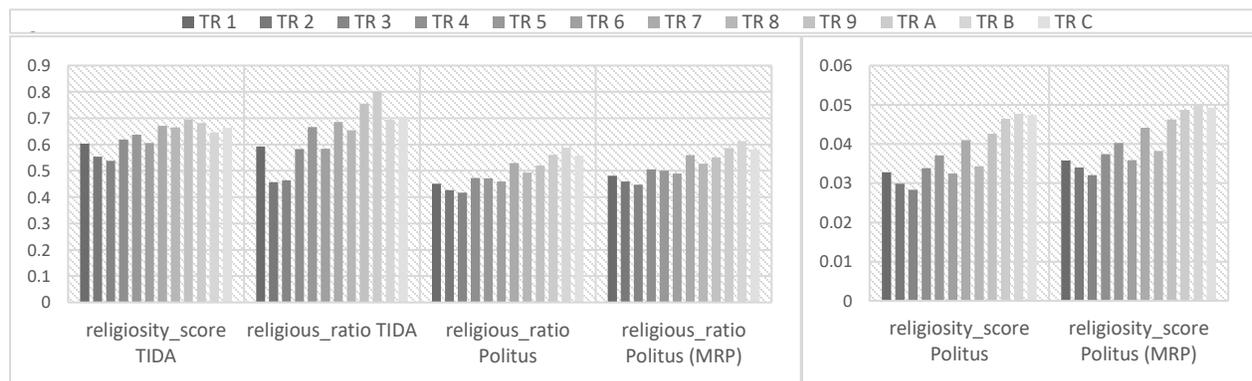

**Figure 3: The Politus vs the TIDA variables**[7]

---

[7] TR 1: İstanbul, TR 2: Batı Marmara, TR 3: Ege, TR 4: Doğu Marmara, TR 5: Batı Anadolu, TR 6: Akdeniz, TR 7: Orta Anadolu, TR 8: Batı Karadeniz, TR 9: Doğu Karadeniz, TR A: Kuzeydoğu Anadolu, TR B: Ortadoğu Anadolu, TR C: Güneydoğu Anadolu.

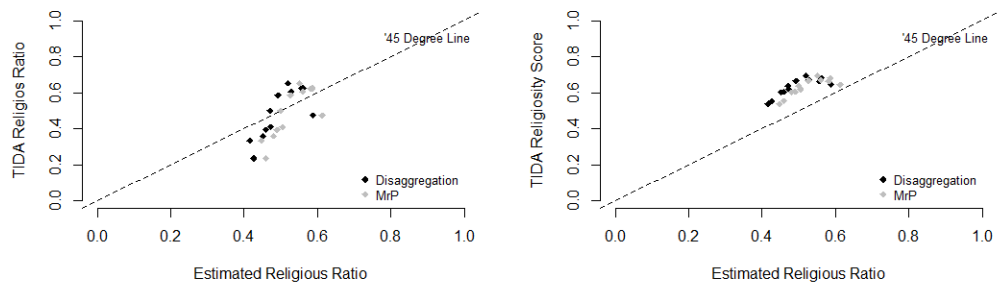

Figure 4: The impact of the MRP analysis for the religious ratio

## ACKNOWLEDGMENTS

This project has received funding from the European Union's Horizon 2020 Research and Innovation programme under Grant Agreement No 952128.

## REFERENCES

[1] Barberá, P., Jost, J. T., Nagler, J., Tucker, J. A., & Bonneau, R. (2015). Tweeting from left to right: Is online political communication more than an echo chamber? Psychological science, 26(10), 1531-1542.

[2] Barberá, P. (2015). Birds of the same feather tweet together: Bayesian ideal point estimation using Twitter data. Political analysis, 23(1), 76-91.

[3] Barberá, P. (2016). Less is more? How demographic sample weights can improve public opinion estimates based on Twitter data. *Work Pap NYU.*

[4] Cohen, R., & Ruths, D. (2013, June). Classifying political orientation on Twitter: It's not easy!. In *Proceedings of the International AAAI Conference on Web and Social Media* (Vol. 7, No. 1).

[5] Filho, Renato Miranda, Jussara M. Almeida, and Gisele L. Pappa. "Twitter population sample bias and its impact on predictive outcomes: a case study on elections." *2015 IEEE/ACM International Conference on Advances in Social Networks Analysis and Mining (ASONAM)*. IEEE, 2015.

[6] Lax, J. R., & Phillips, J. H. (2009). How should we estimate public opinion in the states? *American Journal of Political Science*, *53*(1), 107-121.

[7] Leemann, L., & Wasserfallen, F. (2020). Measuring Attitudes–Multilevel Modeling with Post-Stratification (MrP). The SAGE Handbook of Research Methods in Political Science and International Relations, 371-384.

[8] Olteanu, A., Castillo, C., Diaz, F., & Kıcıman, E. (2019). Social data: Biases, methodological pitfalls, and ethical boundaries. *Frontiers in Big Data*, *2*, 13.

[9] Park, D. K., Gelman, A., & Bafumi, J. (2004). Bayesian multilevel estimation with poststratification: State-level estimates from national polls. *Political Analysis*, *12*(4), 375-385.

[10] Reimers, N., & Gurevych, I. (2019). Sentence-bert: Sentence embeddings using siamese bert-networks. *arXiv preprint arXiv:1908.10084*.

[11] Sen, I., Floeck, F., Weller, K., Weiss, B., & Wagner, C. (2019). A total error framework for digital traces of humans. ArXiv. http://arxiv.org/abs/1907.08228.

[12] Sloan, L. (2017). Who tweets in the United Kingdom? Profiling the Twitter population using the British social attitudes survey 2015. Social Media+ Society, 3(1), 2056305117698981.